# Cs$^+$ Solvated in Hydrogen – Evidence for Several Distinct Solvation Shells

Lorenz Kranabetter,[1] Marcelo Goulart,[1] Abid Aleem,[1,2] Thomas Kurzthaler,[1] Martin Kuhn,[1] Erik Barwa,[1] Michael Renzler,[1] Lukas Grubwieser,[1] Matthias Schwärzler,[1] Alexander Kaiser,[1] Paul Scheier,*,[1] and Olof Echt*,[1,3]

[1] *Institut für Ionenphysik und Angewandte Physik, University of Innsbruck, Technikerstrasse 25, A-6020 Innsbruck, Austria*
[2] *LINAC Project, PINSTECH, P.O. Box Nilore, Islamabad 44000, Pakistan*
[3] *Department of Physics, University of New Hampshire, Durham, New Hampshire 03824, USA*

* Corresponding Authors:
e-mail: paul.scheier@uibk.ac.at
e-mail: olof.echt@unh.edu

**Abstract**
Helium nanodroplets are doped with cesium and molecular hydrogen and subsequently ionized by electrons. Mass spectra reveal H$_x$Cs$^+$ ions that contain as many as 130 hydrogen atoms. Two features in the spectra are striking: First, the abundance of ions with an odd number of hydrogen atoms is very low; the abundance of HCs$^+$ is only 1 % that of H$_2$Cs$^+$. The dominance of even-numbered species is in stark contrast to previous studies of pure or doped hydrogen cluster ions. Second, the abundance of (H$_2$)$_n$Cs$^+$ features anomalies at $n$ = 8, 12, 32, 44, and 52. Guided by previous work on ions solvated in hydrogen and helium we assign the anomalies at $n$ = 12, 32, 44 to the formation of three concentric, solid-like solvation shells of icosahedral symmetry around Cs$^+$. Preliminary density functional theory calculations for $n \leq 14$ are reported as well.

## 1. Introduction

In 1970 Clampitt and Jefferies reported a mass spectrum of (H$_2$)$_n$Li$^+$ cluster ions that were formed by bombarding a frozen H$_2$ target with Li$^+$.[1] (H$_2$)$_6$Li$^+$ was the largest ion that could be identified; the yield of (H$_2$)$_7$Li$^+$ was at least ten times weaker. The authors conjectured that 6 H$_2$ form a solvation shell of octahedral symmetry around the ion. Their conjecture has been confirmed in several theoretical studies.[2-8] Up to six H$_2$ molecules bind in T-shape configuration to Li$^+$. Binding is electrostatic;[9] the H$_2$ bond length remains essentially unchanged, the molecular axes are perpendicular to each other, their distances from Li$^+$ barely depend on $n$, and their dissociation energies decrease gradually from $n$ = 1 to 6 by about 20 to 40 %.[2, 4] The dissociation energy of (H$_2$)$_7$Li$^+$, however, is an order of magnitude smaller than that of (H$_2$)$_6$Li$^+$;[4] the separation of the seventh H$_2$ from Li$^+$ is about twice that of the other six.

Since the report by Clampitt and Jefferies[1] numerous experimental and theoretical studies of metal ions M$^+$ complexed with H$_2$ have been published, partly motivated by the search for materials that can reversibly store hydrogen at high gravimetric and volumetric density.[10-11] The experimental techniques that have been applied include temperature-dependent equilibrium experiments,[12] infrared spectroscopy,[13-14] threshold collision-induced dissociation,[15] and formation of (H$_2$)$_n$M$^+$ in gas-phase collisions.[16]

Most of these studies (recently reviewed by Rodgers and Armentrout)[15] pertain to transition metal ions for which bonding may have covalent contributions.[17] Experimental results for alkali ions (A$^+$) solvated in hydrogen are scarce, and limited to very small complexes. Bowers and coworkers have deduced dissociation energies of (H$_2$)$_n$Na$^+$ and (H$_2$)$_n$K$^+$ ($n$ = 1, 2) from measurements in thermodynamic equilibrium.[12] Bieske and coworkers have reported rotationally resolved infrared spectra of $H_2Li^+$ and $(H_2)_2Li^+$; no distinct features could be identified in the congested and noisy spectrum of $(H_2)_3Li^+$.[18]

We are not aware of any experimental data pertaining to $(H_2)_nA^+$ with $n > 3$, save for the early report by Clampitt and Jefferies.[1] What is the size of solvation shells around alkali ions heavier than $Li^+$? Vitillo et al. have computed the equilibrium distance between the center of mass of $H_2$ and $Li^+$, $Na^+$, $K^+$ and $Rb^+$ with density functional theory (DFT).[9] At the MP2/aug-cc-pVQZ level they obtain values of $R_e$ = 2.01, 2.45, 3.06, and 3.09 Å, respectively (and dissociation energies of $D_0$ = 0.228, 0.118, 0.057 and 0.044 eV, respectively). Thus one would expect a significant increase in the number of $H_2$ in the first solvation shell. Chandrakumar and Ghosh, using DFT, report that $Na^+$ and $K^+$ can accommodate 8 $H_2$ in the first shell.[5-6] They did not explore larger complexes; the solvation shell in $(H_2)_8K^+$ may well be incomplete. Barbatti et al. applied self-consistent-field Hartree-Fock theory and observed that up to 7 $H_2$ are positioned in the first solvation shell of $Na^+$.[4] Adding an eighth $H_2$ led to a geometry with 6 $H_2$ in the first and 2 $H_2$ in the second shell.

In the present work we report a high-resolution mass spectrum of $(H_2)_nCs^+$ with $n$ extending to 65. Cesium was chosen because it is monisotopic and, thanks to its large mass, appears in a mass region that is much less contaminated by background ions than lighter alkalis. The ion yield of $(H_2)_nCs^+$ features a local maximum at $n$ = 8 and abrupt drops at $n$ = 12, 32, 44, and 52. We attribute these features to anomalies in the size dependence of dissociation energies. Extrapolating from the sizes of solvation shells in $(H_2)_nLi^+$ and $(H_2)_nNa^+$ it seems plausible to assign the anomaly at $(H_2)_{12}Cs^+$ to closure of the first, presumably icosahedral solvation shell. The following two anomalies at $n$ = 32 and 44 could then be due to two additional solvation layers of icosahedral symmetry, similar to those in $He_nAr^+$,[19-20] $He_nNa^+$,[21] $(H_2)_nH^-$, and $(D_2)_nD^-$.[22] DFT is used to determine equilibrium geometries of $H_2$ bound to $Cs^+$ by ion-induced dipole forces. Preliminary dissociation energies including harmonic zero-point corrections are reported for $n \leq 14$.

## 2. Experimental and theoretical details

Neutral helium nanodroplets were produced by expanding helium (Messer, purity 99.9999 %) at a stagnation pressure of 20 bar through a 5 μm nozzle, cooled by a closed-cycle refrigerator to 9.42 K, into a vacuum chamber (base pressure about $2 \times 10^{-6}$ Pa). At these temperatures helium nanodroplets contain an average number of $4 \times 10^5$ helium atoms, respectively.[23] The resulting supersonic beam was skimmed by a 0.8 mm conical skimmer and traversed a 20 cm long differentially pumped pick-up chamber containing cesium vapor produced by heating metallic cesium (Sigma Aldrich, purity 99.95 % based on a trace metal analysis) at 60 °C. Hydrogen gas was introduced into the cell at a pressure of about $2 \times 10^{-6}$ mbar.

The beam emerging from the dual pickup cell was collimated and crossed by an electron beam with a nominal energy of 26 eV. The ions were accelerated into the extraction region of a reflectron time-of-flight mass spectrometer (Tofwerk AG, model HTOF) with a mass resolution $m/\Delta m$ = 5000 ($\Delta m$ = full-width-at-half-maximum). The base pressure in the mass spectrometer was $10^{-5}$ Pa. Ions were extracted at 90° into the field-free region of the spectrometer by a pulsed voltage. At the end of the field-free region they entered a two-stage reflectron which reflected them towards a microchannel plate detector operated in single ion counting mode. Additional experimental details have been provided elsewhere.[24]

Mass spectra were evaluated by means of a custom-designed software.[25] The routine includes automatic fitting of a custom peak shape to the mass peaks and subtraction of background by fitting a spline to the background level of the raw data. Cesium is monisotopic, and hydrogen and helium are very nearly so (the natural abundance of deuterium is 0.0115 %, that of $^3$He is 0.000137 %). The abundance of ions is derived by a matrix method.

Equilibrium geometries of H$_2$ molecules bound to Cs$^+$ by ion-induced dipole forces have been estimated using DFT. We used the dispersion corrected ωB97X-D hybrid density functional and the double zeta basis set LANL2DZ including an effective core potential for Cs.[26-28] A similar methodology has been applied previously for (H$_2$)$_n$C$_{60}$Cs$^+$.[29] We report here only the ground state equilibrium geometries and potential energies including harmonic zero-point corrections. A proper treatment of the nuclear quantum dynamics is beyond the scope of this article.

## 3. Experimental Results

Two sections of a mass spectrum are displayed in Fig. 1. Cs$^+$ appears in the left section. The most prominent mass peaks in the right section are due to He$_n$Cs$^+$, $4 \leq n \leq 7$. Mass peaks due to (H$_2$)$_n$Cs$^+$ ($7 \leq n \leq 14$) are prominent as well; a dashed line is drawn to connect their peaks. A local maximum at (H$_2$)$_8$Cs$^+$ and an abrupt drop past (H$_2$)$_{12}$Cs$^+$ are clearly discernible.

(H$_2$)$_{2n}$Cs$^+$ ions are nominally isobaric with He$_n$Cs$^+$; the mass difference between them equals 0.0287 u. The resolution of the instrument is sufficient to fully resolve these ions for $n > 2$.

A remarkable feature in Fig. 1 is the near-absence of H$_{2n+1}$Cs$^+$ ions, i.e. ions containing an odd number of hydrogen atoms. The expected position of these ions is indicated by solid triangles. HCs$^+$ is barely visible; its yield is about 1 % that of H$_2$Cs$^+$. H$_3$Cs$^+$ and H$_5$Cs$^+$ cannot be positively identified. The yield of larger complexes containing an odd number of H atoms is less than 5 % of ions that contain one additional H. In contrast, the relative abundance of H$_{2n+1}$Cs$_2^+$ is large.

A complete mass spectrum showing (H$_2$)$_{2n}$Cs$^+$ ions up to $n = 64$ is provided in the Supplemental Information (Fig. S1). Their abundance is presented in Fig. 2. The distribution is rather flat following the magic 12-mer but there are abrupt drops at $n = 32$, 44, and 52.[30]

Fig. 2 includes the abundance of He$_n$Cs$^+$. For clarity, the abundance was divided by a factor 4, leading to a vertical shift of the distribution on the logarithmic scale. The distribution is declining more quickly than that of (H$_2$)$_n$Cs$^+$; a strong drop appears at $n = 15$. Anomalies at approximately this size have already been reported by other authors.[31-32]

## 4. Discussion

The abundance distribution of (H$_2$)$_n$Cs$^+$ features several anomalies, or magic numbers, namely a maximum at $n = 8$ and abrupt drops at $n = 12$, 32, 44 and 52. We are aware of only one previous experimental study of alkali ions solvated by more than 3 H$_2$ molecules, the one by Clampitt and Jefferies who reported a mass spectrum of (H$_2$)$_n$Li$^+$ ions formed by bombarding a frozen H$_2$ target with Li$^+$.[1] The abundance of the ions gradually declined with increasing size but abruptly dropped beyond $n = 6$. (H$_2$)$_7$Li$^+$ could not be positively identified; the published mass spectrum indicates that its abundance was at least an order of magnitude below that of (H$_2$)$_6$Li$^+$.

The authors' conjecture[1] that the dissociation energy (the energy needed to adiabatically remove the most weakly bound H$_2$) of (H$_2$)$_7$Li$^+$ is much less than that of (H$_2$)$_6$Li$^+$ has been confirmed in several theoretical studies.[4-7, 33] Barbatti et al. report that the dissociation energy drops by about a factor 8.[4]

In the following we summarize noteworthy results from theoretical studies for light alkali ions before contrasting them with our own, preliminary DFT results for (H$_2$)$_n$Cs$^+$: 1) The interaction between the solvent molecules is repulsive hence, for $n \leq 6$, the centers-of-mass of the H$_2$ molecules are distributed over a sphere so as to maximize the distance between them.[4, 7] 2) To minimize the interaction between solvent molecules, the molecular axes of adjacent molecules tend to be perpendicular to each other.[4] 3) The distance between H$_2$ and Li$^+$ remains nearly unchanged for $n \leq 6$.[4, 7] 4) In (H$_2$)$_6$Li$^+$ the H$_2$ molecules are located at the vertices of an octahedron.[4, 7] 5) The H$_2$ are fairly delocalized, especially those outside the first solvation shell.[7]

Observation (1) suggests that incomplete second and higher solvation shells in $(H_2)_nCs^+$ are not likely to exhibit island growth, or facets. Facets appear when heavy, highly polarizable rare atoms grow around an ion; closure of these subshells lead to anomalies in dissociation energies and ion abundances.[34-35]

Thus, anomalies in the present work are likely to indicate the completion of solvation shells. We will first focus on the set of anomalies at $n = 12, 32, 44$. These features have previously been observed in abundance distributions of $He_nAr^+$, $(H_2)_nH^-$ and $(D_2)_nD^-$ and assigned to completion of 3 solvation shells of icosahedral symmetry.[19, 22] The first 12 solvent particles are located at the vertices of an icosahedron, the next 20 at the vertices of a dodecahedron, and the next 12 at the vertices of an icosahedron. A path-integral Monte Carlo (PIMC) study of $He_nAr^+$ by Galli and coworkers has confirmed the interpretation.[20] The solvent atoms form three distinct peaks in the radial distribution function, and the computed dissociation energies exhibit abrupt drops upon completion of each shell.

$He_nAr^+$ (as well as $He_nNa^+$) forms a distinct first solvation layer of icosahedral symmetry because the size of the ion is just "right" given the size of the solvent atoms.[20] Is the size of $Cs^+$ just right if the solvent is $H_2$? The following extrapolation suggests the answer is yes. The first (octahedral) solvation shell of $(H_2)_nLi^+$ is completed at $n = 6$.[2, 4-7] Chandrakumar and Ghosh have used density function theory to study the interaction of $H_2$ with cationic main-group elements; they find that 8 $H_2$ in $S_8$ symmetry complete the first solvation shell of $Na^+$.[5-6] On the other hand, Galli and coworkers observed in their PIMC study that the number of helium atoms in the first solvation shell increases from 12 for $Na^+$ to 18 for $Cs^+$.[21] If we apply that factor (18/12) to the computed number ($n = 8$) of $H_2$ in the solvation shell in $(H_2)_nNa^+$ we predict that 12 $H_2$ will complete a solvation shell around $Cs^+$. So, yes, $Cs^+$ has just the right size to form a complete-shell icosahedral $(H_2)_{12}Cs^+$ and two larger shells of icosahedral symmetry, just as $He_nAr^+$,[19-20] $He_nNa^+$,[21] and, presumably, $(H_2)_nH^-$ and $(D_2)_nD^-$.[22]

Let's now turn to the other two anomalies, at $n = 8$ and 52. The parallel that we have drawn between abundance distributions of $(H_2)_nCs^+$ and $He_nAr^+$ provides no clue: The published distribution of the latter shows no anomaly at $n = 8$ and it did not extend beyond $n = 49$.[19] We have re-inspected our data for larger values of $n$; the distribution is smooth, with very little statistical scatter, for $44 < n \le 200$.

Although highly speculative we note that the preferred structure of a cluster ion with 8 solvent atoms or molecules is, quite often, the square antiprism. This structure appears in recent theoretical studies of $Ar_nNa^+$ and $Xe_nNa^+$ by Spiegelman and coworkers.[35-36] The closed-shell $(H_2)_8Na^+$ also adopts this structure,[5] but for $(H_2)_8Li^+$ the square antiprism seems to be less stable than the octahedral $(H_2)_6Li^+$ with two additional $H_2$ located in a second shell.[4, 7]

Even if the most stable structure of $(H_2)_8Cs^+$ were the square antiprism, why would it feature an enhanced dissociation energy? After all, the first solvation shell can accommodate 4 additional $H_2$. Dissociation energies $D_n$ computed by Barbatti et al. for $(H_2)_nLi^+$ provide a possible clue.[4] $D_5$ is smaller than $D_4$ and $D_6$. The authors attribute this loss in stability to the fact that the $H_2$ molecules in $(H_2)_5Li^+$ cannot orient such that all $H_2$ are orthogonal to adjacent $H_2$. One may speculate that this problem of frustrated orientation is related to the low symmetry of $(H_2)_5A^+$. $(H_2)_7Cs^+$ and $(H_2)_9Cs^+$ are likely to feature low-symmetry ground state structures as well; this might force the $H_2$ molecules into similar energetically unfavorable orientations, resulting in a local maximum for the dissociation energy of $(H_2)_8Cs^+$.

We have performed a DFT study of $(H_2)_nCs^+$ in order to gain more insight into the structure and stability of these ions but the results should be considered preliminary. Locally optimized structures for $(H_2)_nCs^+$ ($n = 1$ to 10) have been found by iteratively adding $H_2$ molecules. For $n = 11$ to 14 we could only find transition states with a few small imaginary frequencies. In agreement with previous studies of lighter alkali ions, $H_2$ molecules prefer to adsorb flat on $Cs^+$ with a preference of approximately 17 meV compared to the radial configuration. The $H_2$ molecules form islands within the first solvation shell although, for low coverage, their mutual distance is larger than in the optimized $(H_2)_2$ dimer. For example, the center of mass distance increases from 3.3 Å for $(H_2)_2$ to 3.81 Å for $(H_2)_2Cs^+$. This increase is likely caused by the induced dipoles. However, we could not find a purely repulsive potential as reported previously for smaller alkali ions.[4, 7] The tendency to grow $H_2$ islands on $Cs^+$ yields less symmetric but more stable structures.

Zero-point corrected dissociation energies $D_0$ are reported in the Supplemental Information (Figure S2). They are between 0.036 eV for $n = 4$ and 0.08 eV for $n = 11$. In this size range they exhibit a slight odd-even oscillation with a preference for odd $n$; the 8-mer does not show enhanced stability, at variance with its enhanced ion abundance. The preference for odd $n$ also contrasts with theoretical work by Barbatti et al. who observed that $(H_2)_4Li^+$ and $(H_2)_6Li^+$ are more stable than $(H_2)_5Li^+$.[4]

The calculated dissociation energy (Figure S2) drops abruptly from 0.061 eV for the 12-mer to almost zero for the 13-mer, consistent with the anomaly in the experimental abundance. The 12-mer shows approximate icosahedral symmetry for the $H_2$ centers with small deviations due to the flat adsorption of the $H_2$ molecules and their mutual orientation. Interestingly, the 13$^{th}$ and 14$^{th}$ $H_2$ molecule are still located in the first solvation shell although their mutual distances are compressed; some distances are smaller than for bare $(H_2)_2$.

Finally we turn to another intriguing feature in our data, the near-absence of $H_{2n+1}Cs^+$. $HCs^+$ is barely visible; its yield is about 1 % that of $H_2Cs^+$. Neither $H_3Cs^+$ nor $H_5Cs^+$ can be positively identified. The preference for ions containing an even number of H atoms contrasts with the preference for ions containing an odd number of H atoms when pure hydrogen clusters are ionized.[37-38] The effect is commonly attributed [39] to the large (1.73 eV) exothermicity of the reaction
$$H_2^+ + H_2 \rightarrow H_3^+ + H. \qquad (1)$$

The energetics will change, of course, in the presence of an ion. Chandrakumar and Ghosh observed that physisorption of $H_2$ on $Na^+$ or $K^+$ is energetically favored over dissociative adsorption.[6] We arrived at the same conclusion for $H_2 + C_{60}^+$.[40] Still, ionization of helium droplets doped with $H_2$ and $C_{60}$ preferentially results in odd-numbered ions, $H_{2n+1}C_{60}^+$; even-numbered ions are disfavored by about a factor two.[40]

What, then, causes the near-absence of $H_{2n+1}Cs^+$ in the present data? One possible mechanism is a difference in the ionization mechanism. Most neutral systems, including $H_2$ and $C_{60}$, the dopant will be solvated in the helium droplet. In this case ionization proceeds via $He^+$ formation and resonant hole hopping, or via highly mobile $He^{*-}$.[41] Formation of $He^+$ requires about 24.6 eV; $He^{*-}$ requires a few eV less. Either way, a large amount of excess energy will be released when the embedded cluster is ionized.

Cesium atoms, however, are highly heliophilic.[42] They will reside in a dimple on the surface of the droplet, even in the presence of a highly polarizable complex such as $C_{60}$ inside the droplet.[43] In this situation, cesium may be ionized via Penning ionization by $He^*$ which is also heliophobic.[44-45] $Cs^+$ will then immerse into the droplet and react with the hydrogen cluster, releasing nothing but the solvation energy of $Cs^+$ in hydrogen, which is small (our value computed for $Cs^+$ solvated in $(H_2)_{12}$ is 0.63 eV). The excess energy would be well below the endothermicity of the reaction
$$H_2Cs^+ \rightarrow HCs^+ + H \qquad (2)$$
for which Kaiser et al have computed a value of 4.75 eV (4.48 eV if harmonic zero-point correction is included).[29] On the other hand, the solvation energy is more than sufficient to cause rapid evaporation of a few solvent molecules, thus leading to the observed anomalies in the ion abundance.

The uncalibrated electron energy used in the present study was 26 eV. However, we barely observe $He_n^+$ in the mass spectrum. This suggests that the actual energy was several eV lower, below the ionization threshold of $He_n$. In this situation direct ionization of a hydrogen cluster via charge transfer with $He^+$ would be excluded. Penning ionization requires only 19.8 eV, thus our experimental conditions favored ionization of cesium on the surface over direct, dissociative ionization of the embedded hydrogen cluster.

There is another important difference between $(H_2)_nCs^+$ on the one hand (which are abundant) and $(H_2)_n^+$ and $(H_2)_nC_{60}^+$ on the other (which are not): The former are electronically closed shell while the latter are open-shell systems. Indeed, we observe closed shell $H_{2n+1}Cs_2^+$ ions with relatively high abundance. Unfortunately this fact does not help to assess the importance of the ionization mechanism because Cs atoms will reside on the helium surface while $Cs_2$ will, in the presence of a polarizable dopant, dive into the droplet.[43] Additional experiments recorded with different electron energies will be required to identify the cause for the near-absence of $Cs^+$ complexed with an odd number of hydrogen atoms.

## 5. Conclusion

Electron ionization of helium droplets doped with cesium and hydrogen results in mass spectra dominated by $(H_2)_nCs^+$ ions; $H_{2n+1}Cs^+$ ions are barely detectable. This is, to the best of our knowledge, the first reported abundance distribution of atomic ions solvated by a large number of $H_2$. The $(H_2)_nCs^+$ series shows anomalies at $n$ = 8, 12, 32, 44, 52. Comparison with previous experiments as well as calculations pertaining to lighter alkali metal ions complexed with $H_2$ and alkali ions solvated in helium suggests that $(H_2)_{12}Cs^+$ represents a $Cs^+$ ion with a complete solvation shell of icosahedral symmetry; this conjecture is confirmed by DFT calculations. The anomalies at $n$ = 32 and 44 are attributed to completion of a second and third solvation shell in which the additional $H_2$ molecules form a dodecahedron and icosahedron, respectively. We have outlined a possible explanation for the anomaly at $n$ = 8 that involves frustrated orientation of the $H_2$ molecules but our preliminary DFT calculations do not reproduce the postulated enhanced stability of the 8-mer. More theoretical work is needed to elucidate the structures of the "magic" 8-mer, 32-mer, 44-mer and 54-mer; future experiments involving $Na^+$, $K^+$ and $Rb^+$ would help to elucidate trends in the solvation of alkali ions in hydrogen.

**Acknowledgements**
This work was supported by the Austrian Science Fund, Wien (FWF Projects I978, M1908, and P26635, P28979).

**Supporting Information**
Fig 1S Mass spectrum ranging from 0 to 263 u.
Fig 2S Computed dissociation energies, including harmonic zero-point corrections, of $(H_2)_nCs^+$ for $n \leq 14$.

**References**
1. Clampitt, R.; Jefferies, D. K., Ion Clusters. *Nature* **1970**, *226*, 141-142.
2. Rao, B. K.; Jena, P., Hydrogen Uptake by an Alkali Metal Ion. *Europhys. Lett.* **1992**, *20*, 307-312.
3. Davy, R.; Skoumbourdis, E.; Kompanchenko, T., Complexation of Hydrogen by Lithium: Structures, Energies and Vibrational Spectra of $Li^+(H_2)_n$ ($n$ = 1 - 4), $Li-H(H_2)_m$ and $Li-H^+(H_2)_m$ ($m$ = 1 - 3). *Mol. Phys.* **1999**, *97*, 1263-1271.
4. Barbatti, M.; Jalbert, G.; Nascimento, M. A. C., The Effects of the Presence of an Alkaline Atomic Cation in a Molecular Hydrogen Environment. *J. Chem. Phys.* **2001**, *114*, 2213-2218.
5. Chandrakumar, K. R. S.; Ghosh, S. K., Electrostatics Driven Interaction of Dihydrogen with s-Block Metal Cations: Theoretical Prediction of Stable $MH_{16}$ Complex. *Chem. Phys. Lett.* **2007**, *447*, 208-214.

6. Chandrakumar, K. R. S.; Ghosh, S. K., Alkali-Metal-Induced Enhancement of Hydrogen Adsorption in $C_{60}$ Fullerene: An Ab Initio Study. *Nano Lett.* **2008**, *8*, 13-19.
7. Ponzi, A.; Marinetti, F.; Gianturco, F. A., Structuring Molecular Hydrogen around Ionic Dopants: $Li^+$ Cations in Small $pH_2$ Clusters. *Phys. Chem. Chem. Phys.* **2009**, *11*, 3868-3874.
8. De Silva, N.; Njegic, B.; Gordon, M. S., Anharmonicity of Weakly Bound $Li^+$-$(H_2)_n$ (*n* = 1 - 3) Complexes. *J. Phys. Chem. A* **2012**, *116*, 12148-12152.
9. Vitillo, J. G.; Damin, A.; Zecchina, A.; Ricchiardi, G., Theoretical Characterization of Dihydrogen Adducts with Alkaline Cations. *J. Chem. Phys.* **2005**, *122*, 114311.
10. Chen, P.; Wu, X.; Lin, J.; Tan, K. L., High $H_2$ Uptake by Alkali-Doped Carbon Nanotubes under Ambient Pressure and Moderate Temperatures. *Science* **1999**, *285*, 91-93.
11. Mavrandonakis, A.; Tylianakis, E.; Stubos, A. K.; Froudakis, G. E., Why Li Doping in MOFs Enhances $H_2$ Storage Capacity? A Multi-Scale Theoretical Study. *J. Phys. Chem. C* **2008**, *112*, 7290-7294.
12. Bushnell, J. E.; Kemper, P. R.; Bowers, M. T., $Na^+/K^+(H_2)_{1,2}$ Clusters - Binding Energies from Theory and Experiment. *J. Phys. Chem.* **1994**, *98*, 2044-2049.
13. Dryza, V.; Poad, B. L. J.; Bieske, E. J., Attaching Molecular Hydrogen to Metal Cations: Perspectives from Gas-Phase Infrared Spectroscopy. *Phys. Chem. Chem. Phys.* **2012**, *14*, 14954-14965.
14. Dryza, V.; Bieske, E. J., Non-Covalent Interactions between Metal Cations and Molecular Hydrogen: Spectroscopic Studies of $M^+$-$H_2$ Complexes. *Int. Rev. Phys. Chem.* **2013**, *32*, 559-587.
15. Rodgers, M. T.; Armentrout, P. B., Cationic Noncovalent Interactions: Energetics and Periodic Trends. *Chem. Rev.* **2016**, *116*, 5642-5687.
16. Dietrich, G.; Dasgupta, K.; Kuznetsov, S.; Lützenkirchen, K.; Schweikhard, L.; Ziegler, J., Chemisorption of Hydrogen on Charged Vanadium Clusters. *Int. J. Mass Spectrom. Ion Proc.* **1996**, *157*, 319-328.
17. Bushnell, J. E.; Kemper, P. R.; van Koppen, P.; Bowers, M. T., Mechanistic and Energetic Details of Adduct Formation and Sigma-Bond Activation in $Zr^+(H_2)_n$ Clusters. *J. Phys. Chem. A* **2001**, *105*, 2216-2224.
18. Emmeluth, C.; Poad, B. L. J.; Thompson, C. D.; Weddle, G. H.; Bieske, E. J., Infrared Spectra of the $Li^+$-$(H_2)_n$ (*n* = 1 - 3) Cation Complexes. *J. Chem. Phys.* **2007**, *126*, 204309.
19. Bartl, P.; Leidlmair, C.; Denifl, S.; Scheier, P.; Echt, O., On the Size and Structure of Helium Snowballs Formed around Charged Atoms and Clusters of Noble Gases. *J. Phys. Chem. A* **2014**, *118*, 8050–8059.
20. Tramonto, F.; Salvestrini, P.; Nava, M.; Galli, D. E., Path Integral Monte Carlo Study Confirms a Highly Ordered Snowball in $^4$He Nanodroplets Doped with an $Ar^+$ Ion. *J. Low Temp. Phys.* **2015**, *180*, 29–36.
21. Galli, D. E.; Ceperley, D. M.; Reatto, L., Path Integral Monte Carlo Study of $^4$He Clusters Doped with Alkali and Alkali-Earth Ions. *J. Phys. Chem. A* **2011**, *115*, 7300-7309.
22. Renzler, M.; Kuhn, M.; Mauracher, A.; Lindinger, A.; Scheier, P.; Ellis, A. M., Anionic Hydrogen Cluster Ions as a New Form of Condensed Hydrogen. *Phys. Rev. Lett.* **2016**, *117*, 273001.
23. Gomez, L. F.; Loginov, E.; Sliter, R.; Vilesov, A. F., Sizes of Large He Droplets. *J. Chem. Phys.* **2011**, *135*, 154201.
24. An der Lan, L.; Bartl, P.; Leidlmair, C.; Schöbel, H.; Jochum, R.; Denifl, S.; Märk, T. D.; Ellis, A. M.; Scheier, P., The Submersion of Sodium Clusters in Helium Nanodroplets: Identification of the Surface → Interior Transition. *J. Chem. Phys.* **2011**, *135*, 044309.

25. Ralser, S.; Postler, J.; Harnisch, M.; Ellis, A. M.; Scheier, P., Extracting Cluster Distributions from Mass Spectra: Isotopefit. *Int. J. Mass Spectrom.* **2015**, *379*, 194-199.
26. Chai, J.-D.; Head-Gordon, M., Long-Range Corrected Hybrid Density Functionals with Damped Atom-Atom Dispersion Corrections. *Phys. Chem. Chem. Phys.* **2008**, *10*, 6615-6620.
27. Dunning Jr., T. H.; Hay, P. J., *Modern Theoretical Chemistry*; vol. 3, Schaefer, H. F., ed., Plenum: New York, 1977.
28. Wadt, W. R.; Hay, P. J., *Ab Initio* Effective Core Potentials for Molecular Calculations. Potentials for Main Group Elements Na to Bi. *J. Chem. Phys.* **1985**, *82*, 284-298.
29. Kaiser, A.; Renzler, M.; Kranabetter, L.; Schwärzler, M.; Parajuli, R.; Echt, O.; Scheier, P., On Enhanced Hydrogen Adsorption on Alkali (Cesium) Doped $C_{60}$ and Effects of the Quantum Nature of the $H_2$ Molecule on Physisorption Energies. *Int. J. Hydrogen Energy* **2017**, *42*, 3078-3086.
30. Ion transmission and detection efficiency of a mass spectrometer are not constant over a wide mass range. However, they will change gradually with mass and not cause any local anomalies in the ion abundance.
31. Müller, S.; Mudrich, M.; Stienkemeier, F., Alkali-Helium Snowball Complexes Formed on Helium Nanodroplets. *J. Chem. Phys.* **2009**, *131*, 044319.
32. Theisen, M.; Lackner, F.; Ernst, W. E., Cs Atoms on Helium Nanodroplets and the Immersion of $Cs^+$ into the Nanodroplet. *J. Chem. Phys.* **2011**, *135*, 074306.
33. Ponzi et al.[7] write that the dissociation energy of $(H_2)_6Li^+$ is about the same as that of $(H_2)_7Li^+$ but a factor 4 weaker than that of $(H_2)_5Li^+$. This implausible conclusion seems to arise from an error in the bottom panel of their Fig. 5. The data in the top panel of Fig. 5 show that the dissociation energy drops between $(H_2)_6Li^+$ and $(H_2)_7Li^+$.
34. Martin, T. P., Shells of Atoms. *Phys. Rep.* **1996**, *273*, 199-241.
35. Slama, M.; Issa, K.; Ben Mohamed, F. E.; Rhouma, M. B.; Spiegelman, F., Structure and Stability of $Na^+Xe_n$ Clusters. *Eur. Phys. J. D* **2016**, *70*, 242.
36. Rhouma, M. B. E.; Calvo, F.; Spiegelman, F., Solvation of $Na^+$ in Argon Clusters. *J. Phys. Chem. A* **2006**, *110*, 5010-5016.
37. van Lumig, A.; Reuss, J., Collisions of Hydrogen Cluster Ions with a Gas Target, at 200-850 eV Energy. *Int. J. Mass Spectrom. Ion Phys.* **1978**, *27*, 197-208.
38. Ekinci, Y.; Knuth, E. L.; Toennies, J. P., A Mass and Time-of-Flight Spectroscopy Study of the Formation of Clusters in Free-Jet Expansions of Normal $D_2$. *J. Chem. Phys.* **2006**, *125*, 133409.
39. Jaksch, S., et al., Formation of Even-Numbered Hydrogen Cluster Cations in Ultracold Helium Droplets. *J. Chem. Phys.* **2008**, *129*, 224306.
40. Kaiser, A.; Leidlmair, C.; Bartl, P.; Zöttl, S.; Denifl, S.; Mauracher, A.; Probst, M.; Scheier, P.; Echt, O., Adsorption of Hydrogen on Neutral and Charged Fullerene: Experiment and Theory. *J. Chem. Phys.* **2013**, *138*, 074311.
41. Mauracher, A.; Daxner, M.; Postler, J.; Huber, S. E.; Denifl, S.; Scheier, P.; Toennies, J. P., Detection of Negative Charge Carriers in Superfluid Helium Droplets: The Metastable Anions $He^{*-}$ and $He_2^{*-}$. *J. Phys. Chem. Lett.* **2014**, *5*, 2444-2449.
42. Ancilotto, F.; Cheng, E.; Cole, M. W.; Toigo, F., The Binding of Alkali Atoms to the Surfaces of Liquid-Helium and Hydrogen. *Z. Phys. B* **1995**, *98*, 323-329.
43. Renzler, M.; Daxner, M.; Kranabetter, L.; Kaiser, A.; Hauser, A. W.; Ernst, W. E.; Lindinger, A.; Zillich, R.; Scheier, P.; Ellis, A. M., Communication: Dopant-Induced Solvation of Alkalis in Liquid Helium Nanodroplets. *J. Chem. Phys.* **2016**, *145*, 181101.
44. Scheidemann, A. A.; Kresin, V. V.; Hess, H., Capture of Lithium by $^4$He Clusters: Surface Adsorption, Penning Ionization, and Formation of $HeLi^+$. *J. Chem. Phys.* **1997**, *107*, 2839-2844.

45. Huber, S. E.; Mauracher, A., On the Properties of Charged and Neutral, Atomic and Molecular Helium Species in Helium Nanodroplets: Interpreting Recent Experiments. *Mol. Phys.* **2014**, *112*, 794-804.

**Figure Captions**

Figure 1
Sections of a mass spectrum of helium nanodroplets doped with cesium and hydrogen. Mass peaks due to $(H_2)_nCs^+$, $7 \leq n \leq 14$, are connected by a dashed line; magic numbers at $n = 8$ and $12$ are indicated. The most prominent peaks are due to $He_nCs^+$ ($4 \leq n \leq 7$). The expected positions of $H_{2n+1}Cs^+$ and $H_{2n+1}^+$ are marked by solid and open triangles, respectively.

Figure 2
Ion abundances of $(H_2)_nCs^+$ and $He_nCs^+$ extracted from the mass spectrum shown in Fig. 1. Prominent anomalies in the distributions are marked.

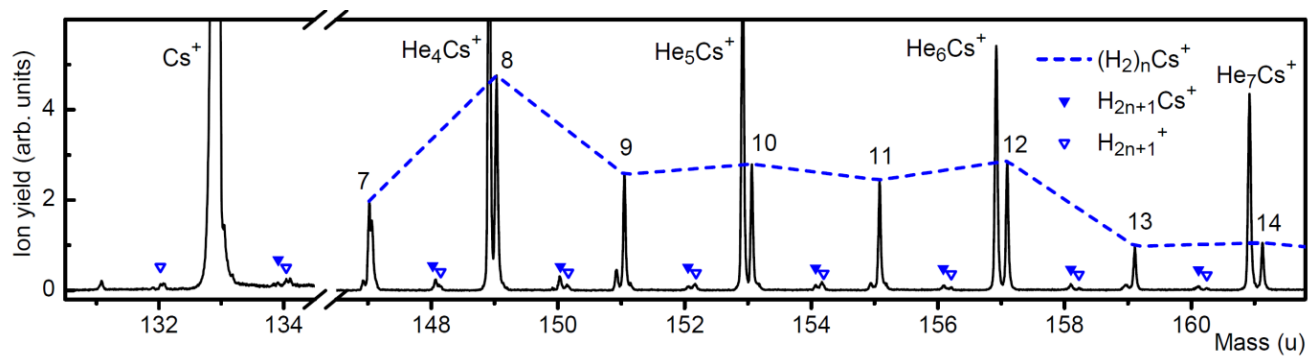

Fig. 1

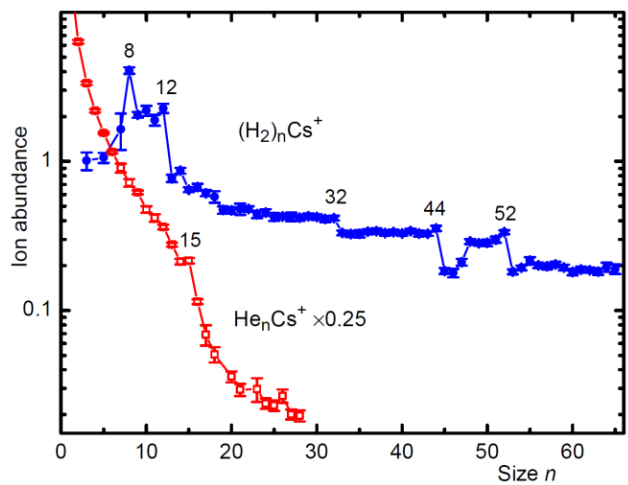

Fig. 2

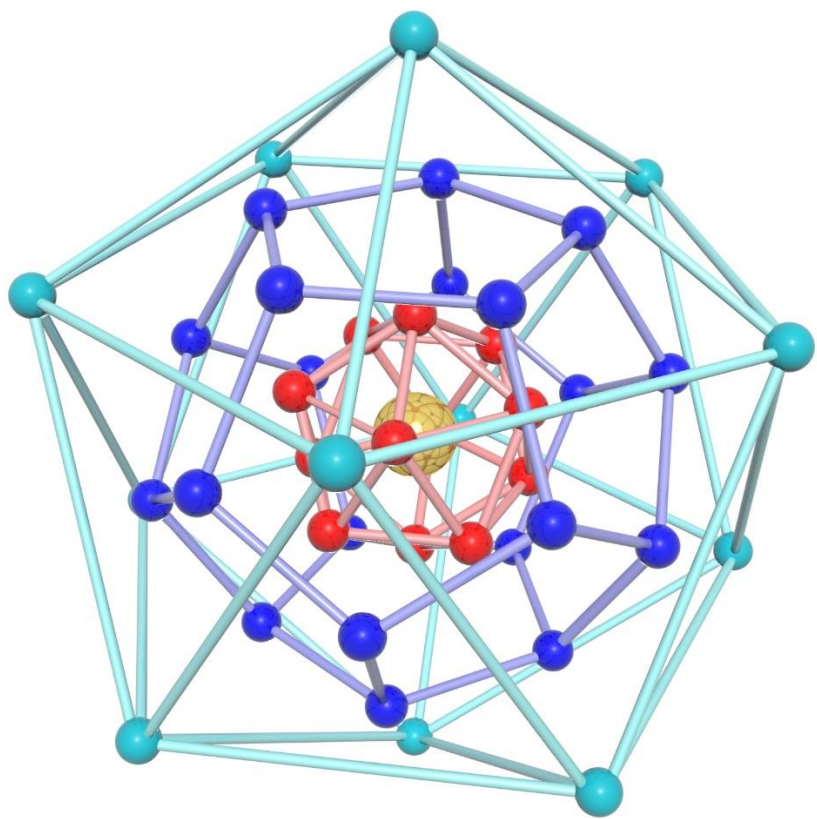

TOC Graphic

**Supporting Information**

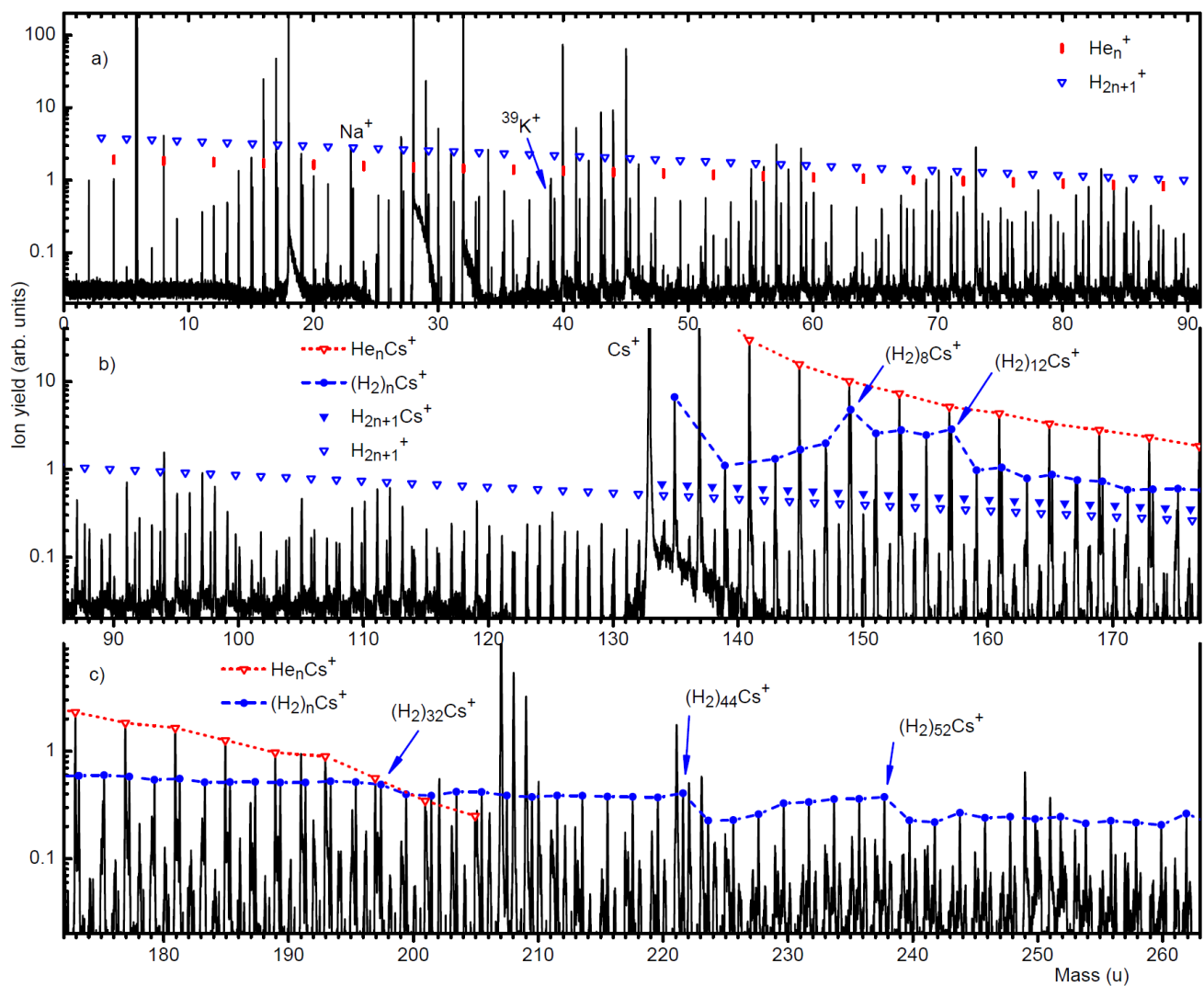

Fig. S1 (Supporting Information)

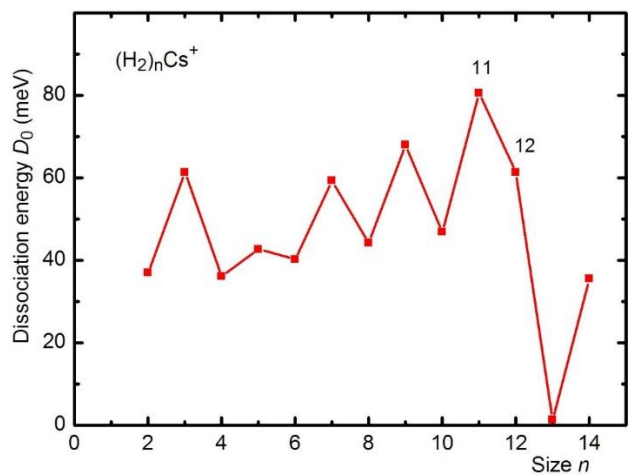

Fig. S2 (Supporting Information)